\documentclass[reprint,twocolumn,amsmath, amssymb, aps, pra]{revtex4-1}
\usepackage{graphicx}
\usepackage{hyperref}
\usepackage{epstopdf}
\usepackage{dcolumn}

\usepackage{bm}

\begin{document}
\title{Suppression of residual amplitude modulation effects in the Pound-Drever-Hall locking}
\author{Xiaohui Shi}
\affiliation{MOE Key Laboratory of Fundamental Physical Quantities Measurement, \\Hubei Key Laboratory of Gravitation and Quantum Physics, \\School of Physics, Huazhong University of Science and Technology, Wuhan 430074, P. R. China}
\author{Jie Zhang}
\email{jie.zhang@mail.hust.edu.cn}
\affiliation{MOE Key Laboratory of Fundamental Physical Quantities Measurement, \\Hubei Key Laboratory of Gravitation and Quantum Physics, \\School of Physics, Huazhong University of Science and Technology, Wuhan 430074, P. R. China}
\author{Xiaoyi Zeng}
\affiliation{MOE Key Laboratory of Fundamental Physical Quantities Measurement, \\Hubei Key Laboratory of Gravitation and Quantum Physics, \\School of Physics, Huazhong University of Science and Technology, Wuhan 430074, P. R. China}
\author{Xiaolong Lv}
\affiliation{MOE Key Laboratory of Fundamental Physical Quantities Measurement, \\Hubei Key Laboratory of Gravitation and Quantum Physics, \\School of Physics, Huazhong University of Science and Technology, Wuhan 430074, P. R. China}
\author{Kui Liu}
\affiliation{MOE Key Laboratory of Fundamental Physical Quantities Measurement, \\Hubei Key Laboratory of Gravitation and Quantum Physics, \\School of Physics, Huazhong University of Science and Technology, Wuhan 430074, P. R. China}
\author{Jing Xi}
\affiliation{MOE Key Laboratory of Fundamental Physical Quantities Measurement, \\Hubei Key Laboratory of Gravitation and Quantum Physics, \\School of Physics, Huazhong University of Science and Technology, Wuhan 430074, P. R. China}
\author{Zehuang Lu}
\affiliation{MOE Key Laboratory of Fundamental Physical Quantities Measurement, \\Hubei Key Laboratory of Gravitation and Quantum Physics, \\School of Physics, Huazhong University of Science and Technology, Wuhan 430074, P. R. China}
\date{\today}
\email{jie.zhang@mail.hust.edu.cn, zehuanglu@mail.hust.edu.cn}
\begin{abstract}
Residual amplitude modulation (RAM) effects in a Pound-Drever-Hall (PDH) technique locked cavity system is analyzed in this paper. The suppression of the amplitude of the RAM has been investigated by many groups, while the effect of the cavity response has not received full attention. Frequency shift caused by RAM in the PDH method is found to be both related to the amplitude of the RAM effects and to the cavity's mode matching and impedance matching. According to our analysis, RAM effects can be fully suppressed by choosing proper impedance matching parameter and magic mode coupling efficiency. We measure the RAM-to-frequency conversion coefficients at different coupling efficiencies. The result agrees well with our theoretical model, demonstrating the potential of full suppression of the RAM effects through proper design of cavities.
\end{abstract}
\maketitle

\section{Introduction}
Ultra-stable lasers have many applications in optical frequency standards \cite{rmp2015_ludlow}, gravitational wave detection \cite{prl2016_abbott} and laser spectroscopy \cite{demtroder2003}.
To improve the stability of a free-running laser, it is often locked to an ultra-stable cavity with the PDH technique \cite{apb1983_drever}.
In the PDH technique, the phase modulation of the laser light is normally performed by an electro-optical modulator (EOM).
Imperfect phase modulation of the EOM can cause residual amplitude modulation (RAM), which in turn can degrade the stability of the locked laser system.
Currently it is believed that the stability of an ultra-stable laser system is going to be dominated by RAM effects at the level of $5\times 10^{-17}$ to $1\times 10^{-16}$ \cite{PRL2017_Matei, JPCS2016_Matei, ol2014_zhang}.

To suppress the RAM effects, a lot of efforts have been paid to minimize the amplitude of the RAM through passive isolation and active suppression for shot-noise-limited detection \cite{ol2014_zhang, josab1985_Hall,josab1985_gehrtz, rsi2012_chen, ol2016_li,ol2016_jiang}.
For passive isolation, wedged EOM crystals, isolators and EOM temperature stabilization have been used \cite{rsi2012_chen,ol2016_li,ol2016_jiang}. At the moment the smallest RAM amplitude is obtained with active feedback control by supplying a servo voltage to the EOM crystal \cite{ol2014_zhang}. According to this upper limit, with a 28 kHz linewidth cavity and $1\times 10^{-6}$ level RAM amplitude control, the stability of ultra-stable lasers at 1070 nm will be limited by RAM at the level of $1\times 10^{-16}$.
A narrower linewidth cavity or a lower RAM amplitude can further improve the laser stability.
This is a great challenge both for coating techniques and RAM suppression. With the same coating mirrors, longer cavity is an easy way to get a narrower cavity linewidth \cite{apb2013_amairi}. However, the vibration noise will increase rapidly at the same time, since the vibration sensitivity for a longer cavity is higher in comparison with that of a shorter cavity \cite{pra2006_chen}. We need to make a compromise between the selection of longer cavity length and further suppression of the RAM amplitude.

To push for better stability, it is worth looking into the effects of RAM on the whole laser system, including the cavity response.
Since the final goal of RAM suppression in PDH locking is to improve the locked laser frequency stability, it is meaningful to investigate the RAM to frequency conversion process. Attention has been seldom paid to this problem before since only an upper bound limit is often used to analyse the RAM effects on laser stability.
In the PDH locking, an offset voltage in the PDH error signal will appear due to RAM effects.
This offset voltage pushes the laser frequency away from the cavity's transmission peak \cite{ajp2001_black, pra2015_chen}.
However, the offset voltage will not contribute any noise when the laser is on resonance with the cavity since the reflection of the carrier will be zero on resonance, which means the laser will be locked to the transmission peak of the cavity tightly even with RAM.

This is not observed in real experiments. It turns out that certain conditions need to be fulfilled in order for the above assumption to be true, which are difficult to be achieved in reality.
The first condition is a pure phase modulation, which is a well understood condition. The second is 100\% transmission through the cavity, which is ruined by imperfect mode matching and impedance matching properties of the cavity. The second condition is difficult to verify due to the fact that the two matching properties are generally combined together and are hard to disentangle if we just measure the cavity's transmission and reflection. We develop a method to measure these two matching properties separately so that the second condition can be verified, and we propose a method for full suppression of the RAM effects using a magic coupling efficiency.

In this paper, we will discuss how RAM affects laser frequency shift with an imperfect laser coupling in Sec. 2. A new theoretical model is deduced considering the mode matching and the impedance matching of the cavity. In Sec. 3, experiments using two ultrastable laser systems are implemented to test the theoretical model and demonstrating the feasibility of further suppression the RAM effects. We conclude in Sec. 4.

\section{Theoretical Analysis}
A phase modulated light with RAM can be written as

\begin{equation}\label{eq:E_PM_RAM_full}
\begin{split}
    E^{PM,RAM}_{inc}(x,y,t)&=E(x,y) e^{i \omega t}[a e^{i(\delta_o \sin\Omega t+\phi_o)}\\
    &+b e^{i(\delta_e \sin\Omega t+\phi_e)}],
    \end{split}
\end{equation}

where $E(x,y)$ is the transverse mode of the injected laser beam, $\delta_{o,e}$ the modulation index, $\Omega$ the modulation frequency and $a=\sin\beta\sin\gamma,\ b = \cos\beta\cos\gamma $ are alignment factors \cite{josab1985_Hall}.
To minimize RAM noise, two high extinction ratio polarizers or isolators are used before and after the EOM, thus, $\beta\approx0,\ \gamma\approx0$ or $\beta\approx \pi/2,\ \gamma\approx \pi/2$, which will result in $a\approx0,\ b\approx1$ or $a\approx1,\ b\approx0$.
Without loss of generality, in the next context, let $a\approx1,\ b\approx0$, which means the extraordinary light makes a perturbation to the ordinary light.
Thus the phase modulation light with RAM can be expanded as
\begin{equation}\label{eq:E_PM_RAM}
\begin{split}
    E^{PM,RAM}_{inc}(x,y,t)\approx E(x,y)e^{i \omega t}e^{i\phi_o}\\
    \times[J_0(\delta_o)+e^{i \epsilon}J_1(\delta_o)e^{i\Omega t} -e^{i \epsilon}J_1(\delta_o)e^{-i\Omega t}],
\end{split}
\end{equation}
where $e^{i \epsilon}=\frac{a J_1(\delta_o)+b e^{i (\phi_e-\phi_o)} J_1(\delta_e)}{J_1(\delta_o)}$ for simplification, and $\epsilon \ll 1$ normally.
The light intensity which incidents into the cavity is $P_{inc}=\iint |E(x,y)|^2\mathrm{d}x\mathrm{d}y$.

If this light is received by a photodiode (PD), the output of the PD is
\begin{equation}\label{eq:V_PM_RAM}
\begin{split}
V(t)&= R\iint \vert E^{PM,RAM}_{inc}(x,y,t)\vert ^2 \mathrm{d}x\mathrm{d}y \\
    &\approx RP_{inc}[1-4\Re(\epsilon) J_0(\delta_o)J_1(\delta_o) \sin(\Omega t)]+[2\Omega\ \textrm{terms}],
\end{split}
\end{equation}
where $R$ is the responsitivity of the PD.
There is a sine term in the Eq. (\ref{eq:V_PM_RAM}) that can be observed in a spectrum analyzer at the modulation frequency.
For pure phase modulation, $\epsilon=0$, $V_{\Omega}(t)=4RP_{inc}\Re(\epsilon) J_0(\delta_o)J_1(\delta_o) \sin(\Omega t)=0$, the signal vanishes.

Usually, RAM signal is captured at the PDH system's demodulation port, and the RAM voltage is:
\begin{equation}\label{V_RAM}
    V_{RAM} = 4\kappa RP_{inc}\Re(\epsilon) J_0(\delta_o)J_1(\delta_o),
\end{equation}
where $\kappa$ is conversion coefficient of the mixer.

As discussed in Sec. 1, 100\% transmission through the cavity is impossible in a real system. Two effects limit the transmission of light through an optical caviyt: one is non-critical impedance matching of the cavity mirrors, and the other is imperfect mode matching between the cavity and the incident light.

To describe the impedance matching of the cavity, the reflection coefficient of a Fabry-Perot cavity can be simplified as
\begin{equation}
    \begin{aligned}
        \label{eq:ref_coef}
        F(\omega)=\frac{E_{ref}}{E_{inc}}&=\frac{r_1 -r_2(r_1^2+t_1^2)e^{-i\omega/\Delta \nu_{fsr}}}{1-r_1r_2e^{-i\omega/\Delta \nu_{fsr}}}\\
        &\approx \frac{\zeta+i\delta\omega/\pi\Delta\nu_c}{1+i\delta\omega/\pi\Delta\nu_c},
    \end{aligned}
\end{equation}
where $r_{1,2}$ is the amplitude reflection coefficient of the two mirrors, $t_1$ is the transmission coefficient of the first mirror, $\Delta\nu_{fsr}=c/2L$ is the free spectral range of the cavity, $\Delta \nu_c$ is the linewidth of the cavity, $\delta \omega = \omega-N\times 2 \pi \nu_{fsr}$ with $N=[\frac{\omega}{2\pi \nu_{fsr}}]$, in addition, $\delta\omega \ll 2\pi\Delta\nu_c$ while the laser is locked to the reference cavity, and $\zeta = \frac{r_1-r_2(r_1^2+t_1^2)}{1-r_1 r_2}$ \cite{ao2007_bondu}.
For critically impedance matched cavity, $\zeta=0$; $0<\zeta\leq1$ is for an undercoupled cavity; and $-1\leq\zeta<0$ is for an overcoupled cavity. For example, for a high finesse cavity a negative impedance match parameter $\zeta$ can be reached if the total loss of mirror 1 is one ppm, $r_1=0.999995$, $r_2=0.9999995$.

Taking mode matching into consideration, we expand the transverse modes of the incident light with the Hermite-Gaussian modes,
\begin{equation}\label{eq:HG_factorization}
    E(x,y)=\sum_{m,n=0}^{m,n=\infty}{E_{mn}(x,y)},
\end{equation}
where $E_{mn}(x,y)$ is the electric field of $\mathrm{TEM}_{mn}$ mode, which is orthogonal to each other.
The cavity's eigen modes $\mathrm{TEM}_{mn}$ are linearly distributed, which is usually much larger than the linewidth of the cavity we used to stabilize the laser.
Thus, only one mode will be on resonant with the cavity, and all the other modes will be totally reflected.
The reflected light can be written as

\begin{equation}\label{eq:E_ref}
\begin{split}
    E_{ref}(x,y,t)=&\sum_{m\neq m_0,n\neq n_0}{E_{mn}(x,y)}e^{i\omega t}e^{i\phi_o}\\
    &\times[J_0(\delta_o)+e^{i\epsilon}J_1(\delta_o)e^{i\Omega t}-e^{i\epsilon} J_{1}(\delta_o)e^{-i\Omega t}] \\
    &+{E_{m_0n_0}(x,y)}e^{i\omega t}e^{i\phi_o}\\
    &\times[F(\omega)J_0(\delta_o)+e^{i\epsilon} J_1(\delta_o)e^{i\Omega t}-e^{i\epsilon} J_{1}(\delta_o)e^{-i\Omega t}],
\end{split}
\end{equation}

where $m=m_0,n=n_0$ is the cavity resonant mode, and the electric field transverse dependence is implicitly implyed.

After detection of this reflected light with a PD, a voltage $V(t) = R\iint{\vert E_{ref}(x,y,t)\vert ^2}\mathrm{d}x\mathrm{d}y$ will be generated.
After some algebra,
\begin{equation}\label{eq:PD_signal}
\begin{aligned}
    V(t)&=R\sum_{m\neq m_0,n\neq n_0} \iint{|E_{mn}(x,y)|^2}\mathrm{d}x\mathrm{d}y \\
    &\times[1-4\epsilon J_0(\delta_o) J_1(\delta_o)\sin(\Omega t)] \\
    &+ R\iint{|E_{m_0n_0}(x,y)|^2}\mathrm{d}x\mathrm{d}y \\
    &\times\vert F(\omega)J_0(\delta_o)+e^{i\epsilon}J_1(\delta_o)e^{i\Omega t}-e^{i\epsilon}J_1(\delta_o)e^{-i\Omega t} \vert^2\\
    &+(2\Omega\ \textrm{terms}).
\end{aligned}
\end{equation}

There is no interference between each TEM$_{mn}$ modes since they are orthogonal to each other. Using a phase-lock method mentioned in Sec. 3, we accurately measure the linewidths of different cavity modes of an ultrastable cavity, and the values are within $3.8$ kHz to $4.5$ kHz for the TEM$_{00}$ to TEM$_{55}$ cavity modes. Therefore we can make a safely assumption that RAM effects for different TEM$_{mn}$ modes are the same. Substitute Eq. (\ref{eq:ref_coef}) into Eq. (\ref{eq:PD_signal}), the $\Omega$ terms in $V(t)$ becomes
\begin{equation}\label{V_err_omega}
\begin{split}
    V_{\Omega}(t) &= 4RJ_0(\delta_o)J_1(\delta_o)P_{inc}\\
    &\times[(1+\zeta\alpha-\alpha)\Re(\epsilon)-\frac{ \alpha(1-\zeta)  \delta\omega}{\pi\Delta\nu_c}]\sin(\Omega t),
\end{split}
\end{equation}

where $\alpha=\frac{\iint{|E_{m_0n_0}(x,y)|^2}\mathrm{d}x\mathrm{d}y}{\iint{|E(x,y)|^2}\mathrm{d}x\mathrm{d}y}$ is the cavity coupling efficiency due to mode matching.
To obtain the error signal, the $\sin(\Omega t)$ signal from the PD is demodulated with a right phase to obtain a maximum error signal, and the error signal is
\begin{equation}\label{eq:error_signal}
\begin{split}
    V_{error}(\delta\omega) &=4\kappa RJ_0(\delta_o)J_1(\delta_o)P_{inc}\\
    &\times[(1+\zeta\alpha-\alpha)\Re(\epsilon)-\frac{ \alpha(1-\zeta)  \delta\omega}{\pi\Delta\nu_c}].
\end{split}
\end{equation}


It is obvious that the error signal in Eq. (\ref{eq:error_signal}) is the sum of an bias voltage due to RAM and a voltage that is proportional to the laser frequency detuning to the cavity's resonance. Let $V_{error}(\delta\omega)=0$ in Eq. (\ref{eq:error_signal}), the locked laser frequency shift from transmission peak due to RAM is
\begin{equation}\label{eq:ram_to_freq}
\begin{split}
  f_{RAM}=\frac{1+\zeta\alpha-\alpha}{2\alpha(1-\zeta)}\Re(\epsilon)\Delta\nu_c.
\end{split}
\end{equation}
This frequency shift contributes to the frequency locking noise.
From Eq. (\ref{eq:ram_to_freq}) we can see that in addition to the relative RAM $\epsilon$, the impedance matching $\zeta$ and mode matching $\alpha$ can also contribute to the frequency noise. For perfect impedance matching ($\zeta=0$) and mode matching ($\alpha=1$), the frequency locking noise due to RAM will be zero even if we have large RAM effects. In another word, the RAM effects will be fully suppressed under this condition.

In order to feasible the experimental test of Eq. (\ref{eq:ram_to_freq}), we rewrite the equation with the following format. The PDH discriminator slope can be obtained from Eq. (\ref{eq:error_signal}) as
\begin{equation}\label{eq:slope}
    D(\alpha) = \frac{V_{error}(\delta\omega)}{\delta\nu} = \frac{8\alpha(1-\zeta) \kappa RJ_0(\delta_o)J_1(\delta_o)P_{inc}}{\Delta\nu_c},
\end{equation}
where $\delta\nu = \delta\omega/2\pi$.
With $\alpha=1$, $\zeta=0$, we can see that the discriminator slope $D_0$ is the same as the one in the Ref. \cite{ajp2001_black}. By substituting Eq. (\ref{V_RAM}) and Eq. (\ref{eq:slope}) into Eq. (\ref{eq:ram_to_freq}), we have

\begin{equation}
\label{eq:ram2freq}
\begin{split}
  f_{RAM}&=\frac{(1+\zeta\alpha-\alpha)V_{RAM}}{8\alpha(1-\zeta) \kappa\textrm{R}J_0(\delta_o)J_1(\delta_o)P_{inc}/\Delta\nu_c}\\
  &=\frac{(1+\zeta\alpha-\alpha)V_{RAM}}{D(\alpha)}.
\end{split}
\end{equation}

Define a RAM to frequency conversion coefficient $\eta=\frac{f_{RAM}}{V_{RAM}/D(\alpha)}$, we then have
\begin{equation}\label{eq:conversion_coef}
\begin{split}
  \eta=1+\zeta\alpha-\alpha.
\end{split}
\end{equation}
To verify Eq. (\ref{eq:ram2freq}), we directly measure the RAM to frequency conversion coefficient $\eta$ under different coupling efficiencies taking advantage of two ultrastable laser systems.

\section{Experimental Results}
\begin{figure}[!htb]
\centering
\includegraphics[width=8cm]{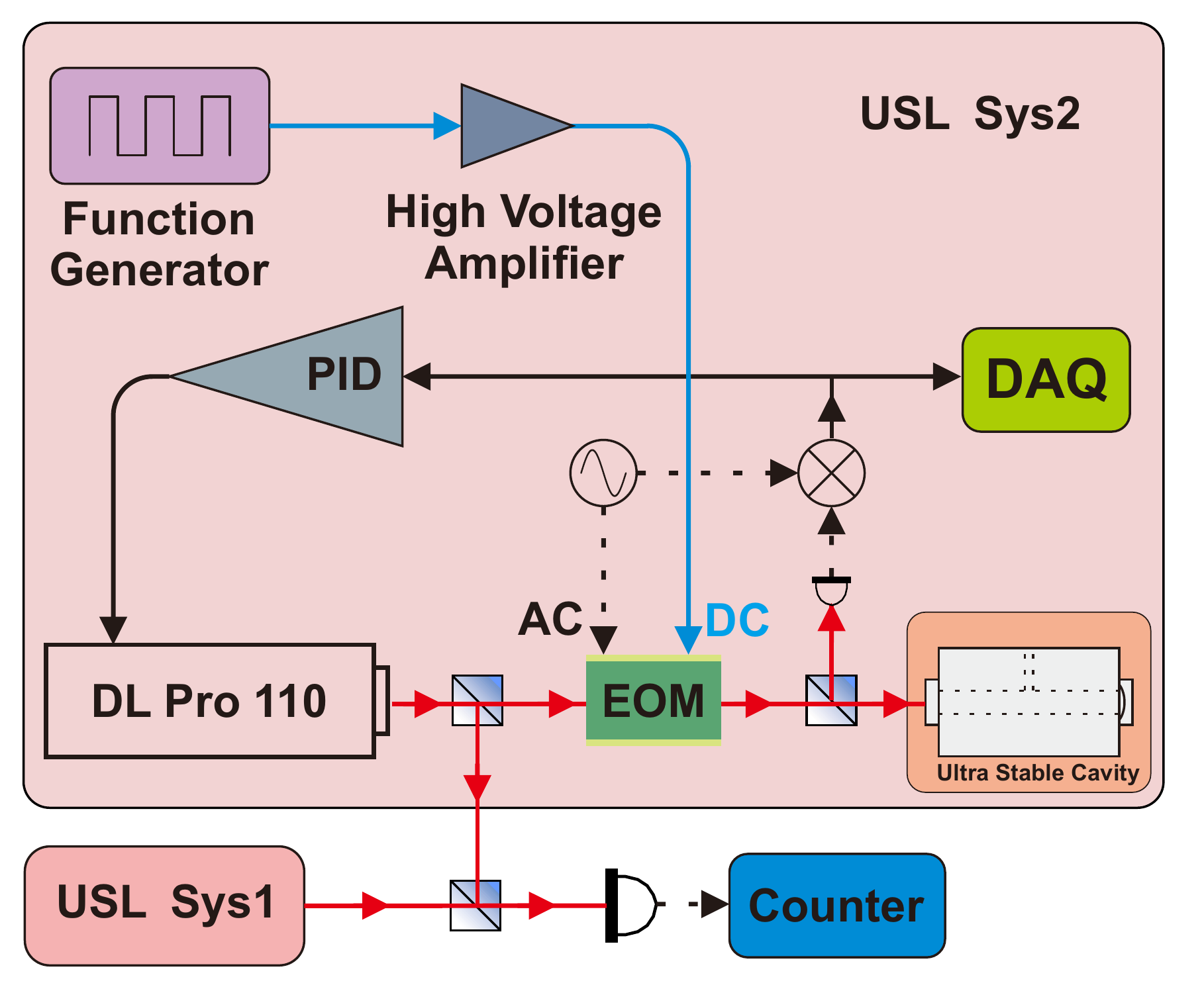}
\caption{Schematic diagram for RAM to frequency noise conversion verification.}
\label{fig_scheme}
\end{figure}

In order to verify Eq. (\ref{eq:ram2freq}) without ambiguity, we can directly measure all the parameters as coupling efficiency $\alpha$, impedance matching parameter $\zeta$, and frequency conversion coefficient $\eta$. We modulate the RAM amplitude and measure the response of the the locked laser frequency at different coupling efficiencies into an ultrastable cavity. To ensure the RAM induced frequency response is large enough to be detected with a high signal to noise ratio (SNR), a high stability laser is needed. In the meanwhile, a similar laser system with the same stability level is needed as a reference to measure the frequency response. As shown in Fig. \ref{fig_scheme}, two ultrastable laser systems (USL Sys1 and USL Sys2) are used here, which are described in detail in Ref. \cite{RSI2016_zhang1}-\cite{RSI2016_zhang2}.
These two ultrastable laser systems are thermal noise limited at a level of $1\times10^{-15}$ and the beat note frequency fluctuation is less than 1 Hz with a 1 s gate time, so the stability level is sufficient for the current experiment. We choose USL Sys1 as a fixed reference laser, and USL Sys2 as a system under test to measure the RAM-to-frequency conversion coefficient. The EOMs (Qubig, EO-T20L3-IR) used in the two systems are covered with well insulated boxes and peltier elements are fixed to the bottom of the EOM for temperature control. The EOM's temperature stability is around 2 mK. In addition, the EOM used in USL Sys2 has a DC modulation port, which is used to modulate the RAM amplitude.

We can obtain the coupling efficiency $\alpha$ by measuring the observed transverse modes from the zeroth to third order modes and calculate the mode-matching ratio of the TEM$_{00}$ mode \cite{ol1995_Uehara,apb1995_Uehara}. But for an ultrastable cavity which has a finesse of $330000$, it is difficult to measure the transmission value accurately, due to a long cavity bulidup time. In order to solve the problem,
instead of locking the laser 2 to the cavity 2 in USL Sys2, we phase lock it to the laser 1 in USL Sys1 with a large tunable offset frequency \cite{pra2013_Hu}. In this way, the laser 2 can be stabilized to a level of $10^{-15}$, and be tuned to resonate with different transverse modes of the cavity 2. We then couple the laser 2 into the cavity 2 and record all the maximum transmission intensities for the different modes. The results are summarized in Table~\ref{table:coupling}, from which we can obtain the coupling efficiency $\alpha$ to the TEM$_{00}$ to be $12.5$ mV$/19.5$ mV$=0.644$. The coupling efficiency $\alpha$ can be tuned over a wide range of values with only two steering mirrors in front of the cavity.

\begin{table}
\caption{Intensity of $\textrm{TEM}_{mn}$ modes from the zeroth to third order modes}
\label{table:coupling}
\centering
    \begin{tabular}{|c|c|c|}
      \hline
      $\textrm{TEM}_{mn}$  &   $V_{peak}$(mV) &$\nu_{qmn}-\nu_{q00} (\textrm{MHz})$\\
      \hline
      00 & 12.5 & 0 \\
      \hline
      01 & 1.6 & 219.332\\
      \hline
      10 & 1.05 & 219.109 \\
      \hline
      02 & 1.5 & 438.704\\
      \hline
      11 & 0.05 & 438.393\\
      \hline
      20  & 1.92 & 438.287\\
      \hline
      03 & 0.52 & 657.940\\
      \hline
      12 & 0.05 & 657.709\\
      \hline
      21 & 0.2 & 657.496\\
      \hline
      30 & 0.1 & 657.462\\
      \hline
    \end{tabular}
\end{table}
Unlike the methods used in Ref. \cite{ao2000_Slagmolen,ao2000_acernese,oe2008_chow}, we use a simpler method to determine cavity 2's impedance matching parameter $\zeta$ under a certain coupling efficiency $\alpha$. We measure the DC voltage difference at the DC port of the PDH error signal photodiode when the laser is locked and unlocked. The difference is
\begin{equation}\label{eq:DC_diff}
    \Delta V = V_{DC, unlocked}-V_{DC, locked}=\alpha(1-\zeta^2)J_0^2(\delta_o)\textrm{R}P_{inc},
\end{equation}
with $V_{DC, unlocked}=\textrm{R}P_{inc}$.
\begin{equation}\label{eq:coupling_efficiency}
    1-\zeta^2 = \frac{\Delta V}{\alpha J_0^2(\delta_o)V_{DC, unlocked}}.
\end{equation}
In our experiment, the phase modulation is adjusted to be $\delta_o=1.08$. $V_{DC, unlocked}$ is measured to be 536 mV and $V_{DC, locked}=371$ mV when the incident power is around $35$ $\mu$W. We then can calculate $|\zeta|=0.32$.

\begin{figure}[!htb]
\centering
\includegraphics[width=9cm]{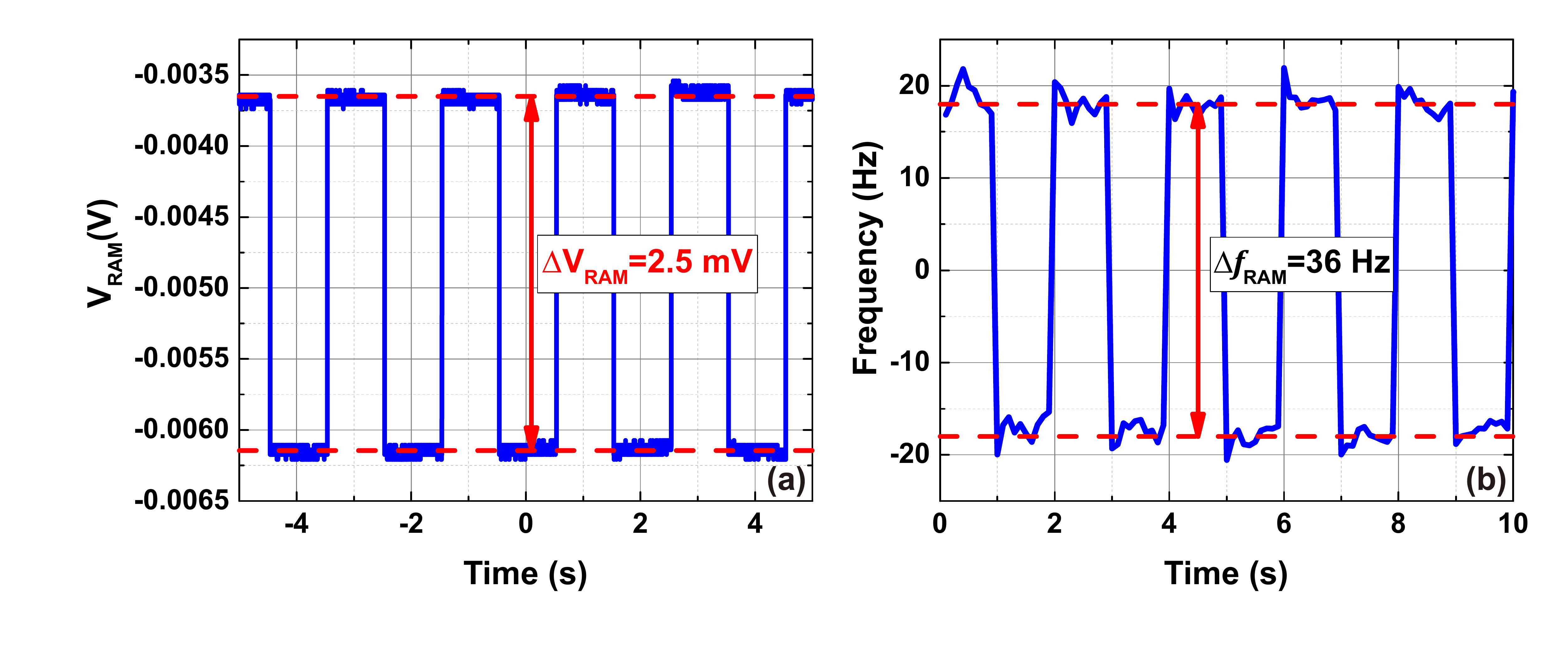}
\caption{(a) Modulated RAM signal with a 150 V square wave signal. (b) Frequency response of the laser at 70\% coupling efficiency with the 150 V modulation.}
\label{fig_RAM}
\end{figure}

To determine a RAM to frequency conversion coefficient $\eta=\frac{f_{RAM}}{V_{RAM}/D(\alpha)}$, we measure $f_{RAM}$, $V_{RAM}$, and $D(\alpha)$, respectively. The discriminator slopes of cavity 2 in USL Sys2 are measured under different coupling efficiencies, and the measurement method has been introduced in Ref. \cite{RSI2016_zhang1}. To modulate the RAM amplitude, a 0.5 Hz, 10 Vpp square wave control voltage is generated by a function generator (Keysight 33522A). To improve the SNR, a PZT amplifier (Thorlabs MDT693B) is used to provide a 15 times amplification. With the 150 V square voltage modulation, RAM signal is measured at the PDH error signal port when the laser is far detuned from the cavity's resonant peak. As shown in Fig. \ref{fig_RAM} (a), the measured RAM voltage $\Delta V_{RAM}$ is 2.5 mV. Although the temperature of the EOM is controlled, the RAM voltage varies with time. We repeat this measurement at different coupling efficiencies.

To obtain $f_{RAM}$, we measure the beat note frequency when USL Sys1 and USL Sys2 are both stabilized, while the USL Sys2 is modulated with the induced 150 V square wave, and USL Sys1 stays as a stable reference. In this way, cavity 2's RAM to frequency response can be determined. Figure \ref{fig_RAM} (b) shows that the RAM induced frequency response has a value of 36 Hz at a 70\% coupling efficiency of cavity 2.

\begin{figure}[!htb]
\centering\includegraphics[width=9cm]{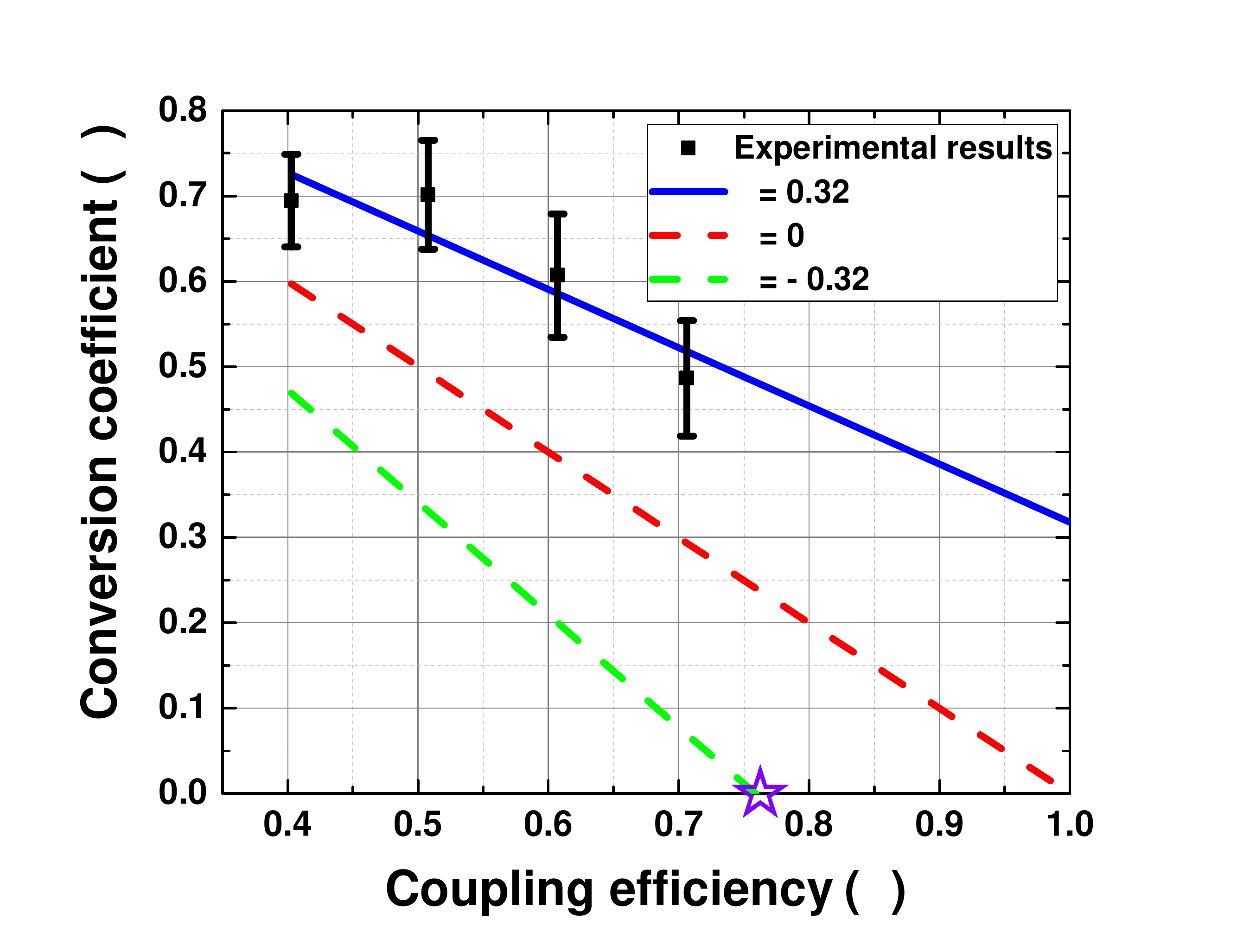}
\caption{RAM to frequency conversion coefficient $\eta$ with respect to coupling efficiency. Red dash curve is the theoretical result according to Eq. (\ref{eq:conversion_coef}) for $\zeta= 0$, blue curve for $\zeta = 0.32$ and green dash curve for $\zeta = -0.32$. A star symbol is used to show the magic coupling efficiency point.}
\label{fig_alpha_ram}
\end{figure}

Frequency responses at different coupling efficiencies are measured so that we can plot the RAM-to-frequency conversion coefficient $\eta$ with respect to coupling efficiency, as shown in Fig. \ref{fig_alpha_ram}. We can see that our experimental results fit well with the theoretical model with $\zeta = 0.32$, which also means that cavity 2 is an under coupled cavity. We can see that in the case of our artificially inflated voltage modulation, for a 70\% coupling efficiency, the RAM induced frequency noise is 36 Hz. With active feedback control of RAM amplitude, RAM voltage fluctuation $\epsilon$ can be suppress to a level of $\epsilon=10^{-6}$ \cite{ol2014_zhang}. Although the USL Sys2's cavity has an impedance matching $\zeta=0.32$, which cannot be easily changed without replacing the cavity mirrors, we can manage a 100\% coupling efficiency. Then the stability limit due to RAM effects can be improved to $5\times10^{-18}$ for a 1070 nm laser. If one pays more attention to manage an optimally coupled cavity $\zeta=0$ and a 100\% coupling efficiency \cite{ol1995_Meuller,oe2010_Rabeling}, the RAM effects will be fully suppressed and have no effect at all to the laser frequency stability even with a big RAM voltage fluctuation $\epsilon$.

More importantly, according to the deduced RAM to frequency conversion relationship in Eq. (\ref{eq:ram2freq}), it is possible to further suppress the RAM effects by using a negative impedance match parameter cavity. As shown with the green dash curve in Fig. \ref{fig_alpha_ram}, for an over-coupled cavity $\zeta=-0.32$, the RAM to frequency conversion coefficient will be much smaller and even reach zero when the coupling efficiency $\alpha=\frac{1}{1-\zeta}=0.76$. We call this coupling efficiency the ``magic coupling efficiency'', shown as a star symbol in Fig. \ref{fig_alpha_ram}. An experiment to test this is ongoing. Note that in our experiment we only discuss RAM effects that are caused by the imperfect phase modulations when lasers pass through the EOM crystal. RAM effects can also be induced by etalons created by multiple reflections between the mirrors and wave plates or by the inhomogeneous spatial distribution in the crystal \cite{josab1985_Hall,rsi2012_chen}. These effects are normally much smaller, and can also be suppressed with the same principle.

\section{Conclusion}
In conclusion, we have developed a formula (Eq. (\ref{eq:ram2freq})) for the RAM to frequency noise conversion taking into account the impedance matching and the mode matching of the cavity in the PDH locking system. With specifically designed experiments, we are able to measure the impedance matching and the mode matching separately and verify the equation without any fitting parameters, and the experimental results fit well with the theoretical calculations. More importantly, ways to totally suppress the RAM induced frequency noise are proposed. One possibility is to design critically coupled cavities and manage a 100$\%$ mode matching; another possibility is to choose an over-coupled cavity and use the magic coupling efficiency. This over-coupling can be achieved by using two different reflectivity mirrors, and coupling light from the the lower reflectivity side into the cavity.
\bibliographystyle{ieeetr}

\section*{Acknowledgments}
We thank Prof. Zhongkun Hu and Prof. Minkang Zhou for the help of laser phase locking. The project is partially supported by the National Key R\&D Program of China (Grant No. 2017YFA0304400), the National Natural Science Foundation of China (Grant Number 91536116, 91336213, and 11774108).

\end{document}